\documentstyle[prb,aps]{revtex}
\begin{document}
\draft


 \title{Coulomb drag in intermediate magnetic fields}

 \author{A.V. Khaetskii}  
\address{ Institute of Microelectronics Technology, Russian
Academy of Sciences,\\
142432, Chernogolovka,
Moscow District, Russia }
\author{Yuli V. Nazarov}
\address{ Faculty of Applied Sciences and DIMES, Delft University of
Technology,\\
Lorentzweg 1, 2628 CJ Delft, The Netherlands}

\date{\today}
\maketitle

\begin{abstract}

We investigated theoretically
the Coulomb drag effect in coupled 2D electron gases in
a wide interval of magnetic field and temperature
$ 1/\tau \ll \omega_c \ll E_F/\hbar$, $T \ll E_F$,
 $\tau$ being intralayer scattering
time,
$\omega_c$ being the cyclotron frequency.
We show that the quantization of
the electron spectrum leads to rich parametric dependences of
drag transresistance on temperature and magnetic field.
This is in contrast to usual resistance. 
New small energy scales are found 
to cut  typical excitation energies to values lower than temperature. 
This may lead to a linear temperature dependence
of  transresistance even in a relatively weak magnetic field and 
can explain some recent experimental data.

We present a novel mechanism of
Coulomb drag when the current in the active layer causes a magnetoplasmon
wind and the magnetoplasmons are absorbed by the electrons of the
passive layer providing a momentum transfer.
We derived  general relations that describe the drag as a result
of resonant tunneling of  magnetoplasmons.

\end{abstract}
\par
\section{Introduction}
When two 2-dimensional electron systems are placed in close proximity,
then
even in the absence of electron tunneling between layers the current in
one
layer (the active layer) will cause the current in the other (the
passive layer). \cite{Gramila}
This phenomenon is known as a
frictional drag and is due to the interlayer Coulomb interaction which
causes
a momentum transfer from one layer to the other.
If no current is allowed  in the passive layer,
the potential difference developes there 
to compensate the frictional interlayer force.
The transresistance is measured as the ratio between 
the electric field developed in
the passive layer to the  current density in the active layer.
This drag is of fundamental interest
because it can be used as a sensitive probe of the screened interlayer
Coulomb
interaction and the form of the irreducible polarizability
function $\chi ( \omega , q)$ within the layer.

Until recently, the experiments on frictional drag have been done
in a
zero magnetic field. \cite{Gramila} In this regime, the theoretical description is 
well elaborated. 
\cite{theory,Flensberg}.
The $T^2$ dependence of  transresistance was explained by the
phase-space arguments.

 A number of experiments on Coulomb drag in  magnetic field have
been also
reported \cite{Hill,Rubel}. Most of them are done in
 Quantum Hall regime
($T \ll \hbar \omega_c \simeq E_F , \;\;
T, \omega_c$ being temperature
and cyclotron frequency respectively) when the screening
of the interaction and the polarizability function are determined by the
states
of the lowermost Landau level.
The experiments have shown a strong filling factor dependence of drag
resistance. It was strongly enhanced (compared to the $H=0$ case) when
the
Fermi level lay within the Landau level and was strongly supressed when
the Fermi
level was between the Landau levels.
 The  recent theoretical work \cite{Bonsager,Bonsager2}
(which treats the case $T < \Delta \ll
\hbar \omega_c \simeq E_F$ , where
$\Delta$ is the width of the Landau level due to disorder) predicts a
twin-peak structure of the transresistance as a function of the 
magnetic field.
This is due to the interplay between the screened interlayer interaction
and the phase-space available for the interlayer e-e scattering.

In the present work, we investigate the
drag effect in intermediate magnetic fields 
$1/\tau \ll \omega_c \ll  E_F/\hbar$. 
In contrast to Quantum Hall regime, the electrons occupy many Landau levels. 
Still these Landau levels 
remain well resolved since their width
$\Delta \simeq \hbar \sqrt{\omega_c /\tau}\ll \hbar \omega_c $ is much
smaller than the level spacing. 

The electron density of states
is strongly distorted in comparison with its zero-field value.
That is why even in the
"classical"
regime $E_F \gg T \gg \hbar \omega_c$ and $R_c \gg d$ ($R_c =
v_F/\omega_c$ being
the cyclotron radius, $d$ being the distance between the layers)
the polarization function $\rm{ Im} \chi(\omega \simeq T , q \simeq 1/d)$ which is
responsible for the
absorbtion of energy, 
differs strongly from its zero-field form. As a function of
frequency it consists of  a
series
of well resolved peaks at multiples of cyclotron frequency.
We will see that this circumstance leads to rich parametric dependences
of transresistance on temperature and magnetic field.
This is to be contrasted with the usual intralayer resistance that exhibits
no anomalies except strongly supressed Shubnikov-de Haas oscillations
and does not manifest the electron density of states.

In the intermediate magnetic field at temperatures $T > \hbar\omega_c$
weakly damped boson excitations become important. 
Those are magnetoplasmons with energies close to multiples
of cyclotron energy.
These excitations provide a new mechanism of
Coulomb drag in the system.
The momentum transfer is provided by  magnetoplasmons being excited in one 
layer and absorbed into the other.
We have found general relations that allow to present  the magnetoplasmon
contribution
to the drag resistance as the result of resonant tunneling of magnetoplasmons. 

The problem we consider in this paper is interesting also
because the experimental dependence \cite{Hill} of transresistance in
the
indicated parameter region has a universal form $\propto B^2 T$.
This temperature dependence is of a highly nontrivial form because the
common
arguments based on consideration of  the phase-space
available for the scattering 
give the $T^2$ dependence. 
We show  that new small
energy scales (for example, $\Delta  \ll T$) play the role of
characteristic cut-off energies which leads to a linear temperature
dependence
of transresistance in a relatively weak magnetic field.

A bird's eye view of our results is provided in figure 1, where we present
temperature and magnetic field exponents of the transresistance
in five distinct parameter regions. We obtain detailed analythical results
for the first three regions that correspond to $T\gg \hbar \omega_c$.
In addition to slow $T-$ and $B-$dependence the transresistance 
in the regions IV-VII ($T\ll \hbar \omega_c$) exhibits oscillatory dependence 
on inverse magnetic field (Shubnikov-de Haas oscillations). 
The oscillations can be hardly
investigated analythically except simplest cases.
In the present paper we only present
analythical results for the regions VI and IV
and give estimations of the transresistance in the regions V and VII
for typical filling factors.

The outline of the paper is as follows. In section 2 we list the
theoretical assumptions we made and present the method which is
essentially the same as in \cite{theory,Flensberg,Bonsager,Bonsager2}.
Details of polarization function and magnetoplasmon spectrum
are presented in section 3. In section 4 we give a detailed description
of magnetoplasmon mechanism of the drag.
A phenomenological description of the resonant tunneling of
magnetoplasmons is elaborated in section 5. We list our analythical
results for high  temperatures in section 6. Section 7 
is devoted to evaluation of the transresistance at low temperatures.

\section{Method}
In the present paper, we cover parameter region 
$\Delta \ll \hbar \omega_c  \ll E_0= \hbar v_F
/d $.
Here $\Delta$ stands for the width of the Landau level.
It  determines the maximun one-particle density of
states in a magnetic field.
The ratio  between $\hbar \omega_c$ and $T$ can be arbitrary.

We assume  here that Landau levels acquire width due to 
scattering by impurities and, following \cite{Ando}, treat the effect in the
selfconsistent
Born approximation (SCBA). This approximation is known to lift
 the
difficulties related to the high Landau level degeneracy.
In this approach, $\Delta^2 = (2/\pi) \hbar \omega_c \hbar/ \tau $.
This expression for $\Delta$  is valid for a
short range
($\delta$-correlated) random potential. 

We also assume Coulomb mechanism of the drag \cite{theory}, so that 
 the d.c. drag current  results
from the  rectification
by the passive layer of the a.c. fluctuating electric field created by
the active one.
In diagrammatic language,
the transconductance is given by a diagramm composed of 
three-body correlation functions, those are connected by Coulomb interaction lines
(photon propagators). It was argued in \cite{Flensberg,Bonsager2} that under very general
conditions the three-body correlation functions can be expressed in terms of electron 
polarization functions $\chi(\omega,{\bf q})_{1,2}$ in each layer.

This yeilds the following expression for the diagonal element of
the transresistivity tensor( Eq. 28 of Ref. \cite{Bonsager2}):
\begin{equation}
\rho^{xx}_{12} =- \frac {\hbar^2}{2e^2} \frac {1}{n_1 n_2 T}
\int \frac{d^2q}{(2\pi)^2}q^2 \int_0^{\infty}\frac{d\omega}{2\pi} 
\mid \frac{V_{12}(q)}{{\cal E} (\omega, q)}  \mid ^2 
\frac{{\rm Im} \chi_1 (\omega,
q)
{\rm Im} \chi_2 (\omega, q)}{\sinh^2 (\hbar \omega/ 2T)}.
\label{2}
\end{equation}
Here $n_1, n_2$ are  electron concentrations in the layers,
$V_{12}(q)$ is the Fourier component of interlayer Coulomb interaction
and ${\cal E}(\omega, q)$ describes (dynamical) screening of this
interaction. Electrostatics gives
$V_{12}=V(q)\exp(-qd), V(q)= 2\pi e^2/ \epsilon q$ ,$\epsilon$ being the
bulk
dielectric constant, and
\begin{equation}
{\cal E} (\omega,q) = (1+ V(q)\chi_1)(1+ V(q)\chi_2) - V_{12}^2 (q)
\chi_1\chi_2.
\label{epsilon}
\end{equation}

It has been argued \cite{Bonsager2} that Eq.1 is valid for arbitrary
magnetic field provided $ q^{-1} \ll l, R_c$, $l$ being mean 
free path.
 Our checks confirm that, so we use
the Eq. \ref{2} in our calculations in intermediate magnetic field regime.

As a reference, we give here the expression for transresistance
in the absence of magnetic field  for temperature $T \ll
E_0$
\cite{theory,Flensberg}:
\begin{equation}
\rho^{xx}_{12}
=\frac{\zeta(3)}{64}
\frac {\hbar}{e^2} \frac {1}{n_1 n_2 d^4} \frac{T^2 a_B^2}
{\hbar^2 v_{F1} v_{F2}},
\label{1}
\end{equation}
Here $\zeta(3) \simeq 1.202 $ is the Riemann zeta- function,
 $v_{F1},
v_{F2}$  are the
corresponding Fermi velocities and $a_B$ is the Bohr radius.

To make use of Eq. \ref{2}, we shall evaluate the polarization
function $\chi(\omega, q; B)$. That we do in the following section.

We will assume that the layers are macroscopically identical.
We keep the indices $1,2$ that label the layers in the formulas
solely for the sake of physical clarity. The exception will be
the discussion of Shubnikov- de Haas oscillations in the regions
IV and VI. We allow there for different filling factors in the
 layers.

\section{Polarization function and magnetoplasmons}

We start with the following  expression for the imaginary part of the
polarization function:
\begin{equation}
{\rm Im} \chi (\omega, q) = \nu \omega_c \sum_{n,m}J_{n-m}^2(qR_c)
\int_{-\infty}
^{+\infty}\frac{d\epsilon}{\pi}[n_F(\epsilon)- n_F(\epsilon + \omega)]
{\rm Im} G_n^r(\epsilon){\rm Im} G_m^r(\epsilon + \omega),
\label{3}
\end{equation}
Here $\nu = m_0/\pi \hbar^2$ is the 2D thermodynamic density of states
in the absence of magnetic field, $n_F$ is the Fermi distribution
function, and $G_n^r$ is
the retarded Green's function of the electrons in the $n$th Landau level.
In the above formula,
we have taken into account  that under the conditions
considered in this paper ($T, \hbar \omega_c \ll E_F$) only big Landau
level numbers are important. Thus the bare vertex function is reduced to its
quasiclassical form, which is  Bessel function of the argument $qR_c$.
In the limit $\Delta \rightarrow 0$ (no disorder) this expression is equivalent
to semiclassical approximations employed in \cite{Aleiner,Old}.

The expression (\ref{3}) disregards vertex corrections due to disorder.
This is safe since we  always assume that $v_F q \gg 1/\tau$ and
$q R_c \gg 1$. We will also disregard the rapidly oscillating
part of Bessel function squares at $q R_c \gg 1$, that is, we assume
that $J_m^2(qR_c) \approx 1/\pi q R_c$. We discuss the relevance of
this assumption in Appendix.

Using the SCBA expression for ${\rm Im} G$ \cite{Ando} in the limit of big $n$,
\begin{equation}
{\rm Im} G_n^r(\epsilon)= -\frac{2}{\Delta}
\sqrt {1- (\frac{\epsilon - \epsilon_n}{\Delta})^2}
\Theta \left (1- (\frac{\epsilon - \epsilon_n}{\Delta})^2 \right),
\label{4}
\end{equation}
where $\Theta$ is the step-function and $\epsilon_n = (n+1/2)\hbar
\omega_c$,we obtain from Eq.(\ref{3}) that
\begin{equation}
{\rm Im} \chi (\omega, q) = \nu \frac{4\omega_c}{\pi^2 \Delta}
\frac{\omega_c}{q v_F}
\sum_{n,m}
\left [n_F(\epsilon_n)- n_F(\epsilon_n + \omega) \right]
X_{im}(\frac{\epsilon_n -\epsilon_m + \hbar \omega}{2\Delta}).
\label{5}
\end{equation}
Here we define dimensionless function
$X_{im}(x) \equiv (4/3)[(1+x^2)E(\sqrt {1-x^2})- 2x^2 F(\sqrt {1-x^2})]$
for $|x|\leq 1$ and $X_{im}(x)= 0$ otherwise. The functions
$F(x), E(x)$ are the complete elliptic integrals of
the first and second kind, respectively. Note that
$\int_{-1}^{+1}dx X_{im}(x)=\pi^2/8. $
The expression for ${\rm Im} \chi (\omega, q)$ assumes  different forms depending
on the temperature. In the most interesting case
 $T \gg \hbar \omega_c$ and $qv_F \gg \omega $ (the latter
inequality is
equivalent to $T \ll E_0$, because the characteristic $q \simeq 1/d$ and
the
frequency cannot be larger than the temperature) we get
from Eq.(\ref{5})
\begin{equation}
{\rm Im} \chi (\omega, q) = \nu \frac{4\omega_c}{\pi^2
\Delta}\frac{\omega}{qv_F}
 \sum_{j=-\infty}^{+\infty}
X_{im}().
\label{6}
\end{equation}
Since $\chi (\omega)$ is an analytical function of $\omega$, 
we can easily obtain the real part from Eq.(\ref{6}).
In the vicinity of the $j$-th cyclotron resonance,
$ |\omega - j\omega_c| \ll \omega_c $, it reads:
\begin{equation}
Re \chi (\omega, q) =\nu + \nu \frac{4\omega_c}{\pi^2
\Delta}\frac{\omega}{qv_F}
X_{re}(\frac{\omega -j\omega_c }{2\Delta}); 
\;\; X_{re}(x)= \frac{1}{\pi}{\rm v.p.}\int_{-1}^{+1}\frac{dy X_{im}(y)}{(y-
x)}.
\label{7}
\end{equation}
The functions $X_{re}(x)$ and $X_{im}(x)$ are plotted in Fig.2.
\par

In the opposite limit of $\Delta \ll T \ll \hbar \omega_c$
and $\omega \leq 2\Delta ; qR_c \gg 1$
we obtain from Eq.(\ref{5}):
\begin{equation}
{\rm Im} \chi (\omega, q) = \nu \frac{4\omega_c}{\pi^2 T}\frac{\omega}{\Delta}
X_{im}( \frac{\omega }{2\Delta}) \frac{1}{qR_c}f_n (1-f_n),
\label{8}
\end{equation}
where $f_n = 1/(1+ \exp(\epsilon_n -\mu) /T)$ being the filling factor
of the n-th Landau level in the layer, $\mu$ being chemical potential.

Finally, at $T\ll \Delta$ we can set $n=m$ and integrate expression
(\ref{3}) over $\epsilon$ in close vicinity of $\mu$. This gives
\begin{equation}
{\rm Im}\chi =\frac{4 \nu \omega_c^2}{\pi^2 \Delta^2} 
\frac{\omega}{q v_F} (1-\frac{(\mu - \epsilon_n)^2}{\Delta^2})
\label{nultempchi}
\end{equation}

Although the magnetoplasmon modes of a 2DEG have been extensively studied 
\cite{Magnetoplasmons}, very little attention has been paid to
their properties at high frequencies in short-wave limit.
Since those are of interest for us, we investigated them in some
detail.  
The dispersion curves of the magnetoplasmon modes in the case of
the weak
damping are determined by the equation $Re {\cal E} = 0$, 
${\cal E}$ being
given by Eq.\ref{epsilon}.
From this equation we get $(Re \chi /\nu) + z_{\pm} = 0$, where
$z_{\pm}= qa_B/2(1 \pm \exp(-qd)) \ll 1$. Using Eq.(\ref{7}) for $Re
\chi$, for each $j$ we obtain two solutions corresponding to
two ($\pm$) magnetoplasmon modes. First thing to note is that 
under our assumptions $qa_B \ll 1$ so that on a large frequency scale
$z_{\pm}$ can be safely omitted. 
The resulting equation $Re \chi =0$ determines a series of Burstein-like
magnetoplasmons \cite{Old} with frequencies that in the interesting region of $q$ are close
to cyclotron harmonics $j \omega_c$. (see Fig. 3) If $q < q_0 \equiv -4 j X_{re}(1)\omega_c^2/\pi^2 \Delta v_F 
\simeq 0.19 j\omega_c/R_c \Delta$ the root of
dispersion relation lies beyond the electron adsorption bands and the
magnetoplasmon is not damped. These magnetoplasmons are of no interest
for us since they do not talk to electrons and thus cannot participate
in drag. If $q \gg q_0$ magnetoplasmons lie deep in the electron
adsorption band and are strongly damped, so that their conribution
to drag cannot be distinguished from the contribution of
electron-electron scattering. This is why we concentrate now 
on a close vicinity of $q_0$ (right panel of Fig. 3).

In this vicinity $\chi(\omega,q)$ can be expanded in Taylor series 
in terms of 
$w = \omega-j\omega_c -2\Delta$, $\kappa=q-q_0$, assuming that $w \ll
\Delta, \kappa \ll q_0$,
\begin{equation}
\chi/\nu = C_2 \frac{w}{2\Delta} +\frac{\kappa}{q_0} +
i C_1 (\frac{w}{2\Delta})^2 \Theta(-w)
\label{chionedge}
\end{equation}
Here $C_{1,2}$ are numerical constants characterizing behaviour
of $X_{im}, X_{re}$ near $x=1$, $C_1 \simeq 6.23$, $C_2 \simeq 2.19$.
This determines the dispersion law of magnetoplasmons
\begin{equation}
w_{\pm} = -\frac{2\Delta}{C_2}( z_{\pm} + \kappa/q_0)
\end{equation}
and their damping
\begin{equation}
\Gamma = \Theta(-w) \frac{C_1 w^2}{C_2  \Delta}
\label{gamma}
\end{equation}
where $z$ is taken at $q=q_0$.
Symmetric and asymmetric modes are split by
\begin{equation}
\delta\omega = \Delta \frac{q_0 a_B}{C_2 \sinh q_0 d}.
\label{deltaw}
\end{equation}
We see that $ \Gamma \ll w \ll \Delta$,
$\delta \omega \ll \Delta$, $\delta \omega$ can be comparable 
with $\Gamma$.

\section{ Magnetoplasmon contribution.}

In this section we consider the magnetoplasmon mechanism of Coulomb
drag.  

In the absence of magnetic field  there are two plasmon modes in 
double-layer system, the one with the electron densities in the two layers
oscillating in phase (the optic mode), and the other one where 
the oscillations are out of
phase (the acoustic mode) \cite{Sarma}.
It was pointed out in \cite{Flensberg} that the drag effect can be greatly
enhanced by dynamical "antiscreening" of the interlayer interaction due to coupled plasmon modes. 
Since the plasmon modes lie beyond
the $T=0$ particle-hole continuum, temperatures of the order of Fermi energy
are required for  a large plasmon
enhancement of the drag effect. 
Only then the thermally excited electrons and holes  with
plasmon velocities provide sufficient damping of the plasmon modes
and thus facilitate plasmon interaction with electrons\cite{Flensberg}.

In the revised situation of intermediate magnetic field,
 the magnetoplasmons have even better chances
to enhance the drag. First, there are many modes and their
typical energies are of the order of $\hbar \omega_c$. 
Therefore these modes can be excited at temperatures  much
lower than Fermi energy. Second, the magnetoplasmons 
in our model acquire natural damping: due to finite
Landau level width they may
lay within the particle-hole continium.
The finite temperature without disorder does not lead to
 magnetoplasmon damping and  to the drag effect. This is in  
contrast to the situation without magnetic field, where ${\rm Im} \chi$ at the plasmon
frequency was calculated for collisionless plasma \cite{Flensberg}.

The magnetoplasmon mechanism of the Coulomb drag, when the current in
the active layer causes a magnetoplasmon wind and the magnetoplasmons are
absorbed by the electrons of the passive layer leading to transfer of
the momentum, must be quite general one. In this section, we evaluate 
magnetoplasmon contribution with using Eq. \ref{2}. 
It turns out that the answer
can be expressed  through only two quantities for each double plasmon
mode:
frequency splitting $\delta \omega$
 and damping $\Gamma$. Any concrete model would only set 
specific expressions for 
$\Gamma$ and $\delta \omega$.
This clarifies the physical meaning of  Eq. \ref{2}. 

To prove this, we rederive the result in the next section in a
phenomenological framework.
 
Let us expand $\chi(\omega,q)$ around the frequency $\omega(q)$
at which ${\rm Re} \chi=0$, 
\begin{equation}
\chi(w,q)= \chi' w + i \chi'',
\end{equation}  
$w$ being frequency deviation.
Here $\chi' = {\rm Re} \ d\chi/d \omega(\omega(q))$, $\chi''={\rm Im} \chi(\omega(q))$.
Expression for ${\cal E}$ reduces to the form
\begin{equation}
{\cal E}=(V^2(q)-V_{12}^2(q))(w \chi' +i \chi'' +\nu z_+)(w \chi' +i \chi'' +\nu z_-)
\label{epsexp}
\end{equation}
Consequently, the integrand in Eq. \ref{2} 
has a sharp maximum near $\omega(q)$ as a function of $\omega$.
This suggests we can now integrate over $w$ in infinite limits.

Eq. \ref{epsexp} determines magnetoplasmon spectrum and 
suggests that mode splitting $\delta \omega=\nu (z_- -z_+)/\chi'$, damping  
$\Gamma=2 \chi''/\chi'$. All factors $V_{12}$, $V$ can be absorbed into these
two quantities. Typical $w$ that contribute to the integral
are of the order $\max (\Gamma, \delta \omega)$. The approximation
is valid if $\chi$ and statistical factor $\sinh^2(\omega/2T)$ 
do not change much in this frequency window,
that is $\omega(q) \gg \delta \omega, d \Gamma/d \omega \ll 1$,
$\max (\Gamma,\delta \omega) \ll T/\hbar$.

Provided these conditions are fulfilled,
we can reduce the expression for the transresistance to the following
elegant
form:
\begin{equation}
\rho^{xx}_{12}=
- \frac {\hbar}{16e^2 n_1 n_2 T }
\int \frac{d^2q q^2}{(2\pi)^2} \sum_{modes}
\frac{1}{\sinh^2(\hbar
\omega(q)/2T)}
\frac{(\delta \omega)^2 \Gamma}{[(\delta \omega)^2 + \Gamma ^2]}.
\label{14}
\end{equation}

Summation over modes means summation over all possible roots
of ${\rm Re} \chi=0$.

\section{Resonant tunneling of magnetoplasmons}

We give here another derivation of this formula which clarifies
its physical meaning.
To describe resonant tunneling of plasmons between the layers,
we introduce for each plasmon mode
a density matrix $\rho_{ij} = <b^{\dagger}_i b_j>$.
Here $i,j=1,2$ label the layers, $b^{\dagger}, b$ are boson 
creation/annihilation operators.
In the absence of dissipation, that is, plasmon emission and
absorption, the density matrix obeys the equation
\begin{equation}
\frac{\partial \rho_{ij}}{\partial t} = Ä i\sum_l
(H_{il}\rho_{lj}-\rho_{il}H_{lj})
\end{equation}
For identical layers, diagonal elements of the Hamiltonian are equal
to each other and dissappear from the equation. The non-diagonal element
that
is responsible for plasmon tunneling between layers can be readily
express in terms of splitting $\delta\omega$ between symmetric and
assymmetric
plasmon state: $H_{12}=H_{21}=\delta\omega/2$.

 The dissipation takes place independently in each of the layers.
It contributes to the time derivative of
the diagonal density matrix elements in the following way:
\begin{equation}
(\frac{\partial \rho_{ii}}{\partial t})_{diss}= \Gamma_i (n^B_i
-\rho_{ii}),
\end{equation}
two terms corresponding to generation and absorption of the plasmons.
The temperature of Bose distribution function $n^B_i$ corresponds to
the electron temperature of the $i$th layer. Non-diagonal matrix
elements
aquire a damping equally from both layers
\begin{equation}
(\frac{\partial \rho_{ij}}{\partial t})_{diss}= - \frac{(\Gamma_i
+\Gamma_j)}{2}
\rho_{ij}
\end{equation}

The system of equations that incorporates both dissipation
and resonant tunneling reads as follows:
\begin{eqnarray}
\frac{\partial \rho_{11}}{\partial t}=\Gamma_1 (n^B_1 -\rho_{11})+
i\frac{\delta \omega}{2}(\rho_{12}-\rho_{21}) \nonumber \\
\frac{\partial \rho_{22}}{\partial t}=\Gamma_2 (n^B_2 -\rho_{22})+
i\frac{\delta \omega}{2}(\rho_{21}-\rho_{12}) \nonumber \\
\frac{\partial \rho_{12}}{\partial t}=
-\frac{(\Gamma_1+\Gamma_2)}{2}\rho_{12}+
i\frac{\delta \omega}{2}(\rho_{11}-\rho_{22}) \nonumber
\end{eqnarray}
where $\rho_{21}=\rho_{12}^*$.
The stationary solution takes the form
\begin{equation}
\rho_{11}= n^B_1 - \frac{\Gamma_2}{\Gamma_1+\Gamma_2} \frac{\delta \omega^2
(n^B_1 -n^B_2)}{\delta \omega^2 +\Gamma_1 \Gamma_2}
\end{equation}
Expression for $\rho_{22}$ is obtained by reverting indices $1,2$.

We can now evaluate the drag force acting on electrons of each layer
by equating it to a momentum flow between the layers.
We sum over modes with all possible $q$ and obtain
\begin{equation}
{\bf F}= - \sum_{\bf q} \hbar {\bf q} ( \frac{\partial
\rho_{ii}}{\partial
t})_{diss} = - \sum_{\bf q} \hbar {\bf q} \frac{\Gamma_1
\Gamma_2}{\Gamma_1+\Gamma_2} \frac{\delta \omega^2
(n^B_1 -n^B_2)}{\delta \omega^2 +\Gamma_1 \Gamma_2}
\label{flow}
\end{equation}
where $\Gamma,n^B,\delta \omega$ may be q-dependent.

We assume that the current flows in the layer $2$ and the drag force
in the layer $1$ is equilibrated by the electric field. The
transrestivity is essentially the ratio of this field to that current.
The effect of the current is that the $n^B_2$ is the equilibrium Bose
distribution in the reference frame where the electrons of the second
layer are in average at rest, rather than in the laboratory reference
frame.
So that
\begin{equation}
n^B_2(\epsilon({\bf q}))= f_B(\epsilon({\bf q})- \hbar ({\bf
v}_{drift}{\bf q}))
\approx f_B(\epsilon({\bf q})) - \hbar ({\bf v}_{drift}{\bf q})
\frac{\partial f_B}{\partial \epsilon}
\label{n2}
\end{equation}
${\bf v}_{drift}$ being the drift velocity.

Substituting (\ref{n2}) in (\ref{flow}) we obtain
\begin{equation}
{\bf F}= \hbar^2 {\bf v}_{drift} \int \frac{d^2 q q^2}{8 \pi^2}
\frac{\Gamma_1 \Gamma_2}{\Gamma_1+\Gamma_2} \frac{\delta \omega^2
}{\delta \omega^2 +\Gamma_1 \Gamma_2}
\frac{\partial f_B(\epsilon({\bf q}))}
{\partial {\epsilon}}
\end{equation}
The last things to note  are that $F=e n_1 E$ and $I=e n_2 v_{drift}$.
If we use this  and set $\Gamma_1=\Gamma_2=\Gamma$ we reproduce
Eq.(\ref{14}).
\par

\section{Results: high temperatures}
\par

In this section we consider the drag resistance at 
 temperatures $T \gg \hbar \omega_c$, which are sufficiently high
to excite  magnetoplasmons and electrons
in many Landau levels. As we can see in Fig. 1,
at high temperature we encounter at least three distinct regions
with different temperature and magnetic field exponents. 

First of all, we shall explain why there are so many regions.
If we compare imaginary part of polarization function with and
without magnetic field, we see that their values averaged over
frequency intervals bigger than $\hbar \omega_c$ are the same
$<({\rm Im}\chi)> = <({\rm Im}\chi(H=0))>$. However, ${\rm Im}\chi =0$ beyond
narrow adsorption bands (Fig. 3). This means that whinin the
bands ${\rm Im} \chi$ is significantly enhanced in comparison with
its zero-field value, typically by a factor $\omega_c/\Delta$.  
Surprisingly enough, the enhancement of ${\rm Im} \chi$ can lead both to
enhancement and suppresion of the drag.

If ${\rm Im} \chi \ll \nu$, ${\rm Re} \chi \approx \nu$. The denominator
${\cal E}$ in Eq. \ref{2} which is responsible for screening
of interlayer potential is the same as without magnetic field.
We refer to this situation as to {\it normal screening} regime.
In this regime, the transresistance  is enhanced in comparison 
to its value without magnetic field, since the effect is
proportional to $<({\rm Im}\chi)^2> \gg <({\rm Im}\chi)>^2 \simeq <({\rm Im}\chi(H=0))>^2$.

At further increase of ${\rm Im} \chi$,  ${\rm Re} \chi$  develops as well so that
both $\mid {\rm Re} \chi \mid, {\rm Im} \chi$ become bigger than $\nu$.  
The denominator $\cal E$  strongly increases. 
That efficiently screens out the inter-layer interaction and leads to drastic descrease
of the transresistance. That we will call {\it overscreening}.

However, $\cal E$ can also decrease with increasing ${\rm Im} \chi$ and
pass zero. Near this line, the inter-layer interaction is greatly
increased.  This is where the {\it magnetoplasmon} contribution
dominates.

The actual value of the drag effect is thus determined by
interplay of these three competing tendencies. 

Let us first evaluate the magnetoplasmon contribution.
Substituting expressions (\ref{gamma}), (\ref{deltaw}) to Eq.(\ref{14}) 
we notice that the integrand  has a sharp extremum near
$\kappa \approx 0$
so we can formally integrate over $\kappa$ in infinite limits.
This yields the relation which is valid for all regions, 
\begin{equation}
\rho^{xx}_{12}=
- 0.00221 \frac {\hbar \Delta}{ e^2 n_1 n_2 d^4 T }
\left (\frac{a_B}{d} \right )^{3/2}\alpha^4
\sum_{j=1}^{\infty} \frac{(\hbar \omega_j/T)^4}{\sinh^2(\hbar \omega_j
/2T)}
\left (\frac{\alpha \hbar \omega_j/T}
{\sinh (\alpha \hbar \omega_j/T)} \right )^{3/2},
\label{15}
\end{equation}
where $\alpha \simeq 0.19 (T\omega_c/\Delta E_0)$.

This expression can be simplified further.
The region I ($\hbar \omega_c \ll T \ll E_0 \Delta /\hbar \omega_c$)
corresponds to $\alpha \ll 1$.
Since $\hbar \omega_c \ll T$ 
the characteristic values of $j$ in the sum (\ref{15})
are much bigger than unity. Therefore we can convert the sum over $j$ into the integral.
Since
$\int_0^{\infty} dx x^4 \sinh^{-2} (x/2)= 16 \pi^4/15$, we obtain
\begin{equation}
\rho^{xx}_{12} \simeq
- 0.0003 \frac {\hbar }{e^2 n_1 n_2 d^4}
\frac{(\hbar \omega_c)^3 T^4}{\Delta^3 E_0^4}
\left (\frac{a_B}{d} \right )^{3/2}.
\label{16}
\end{equation}

The region II is defined by inequalities $T \gg E_0 \Delta /\omega_c \gg \omega_c$.
 Here $\alpha \gg 1$. As a result, the
characteristic frequencies here $\tilde \omega_j \simeq E_0 \Delta
/\omega_c \ll
T$ (see Eq.(\ref{15})).  These frequencies, however, are still much
larger than
cyclotron frequency: $\tilde \omega_j \gg \omega_c$.
This enables us again to introduce the continuous variable $x= \alpha
\hbar \omega_j /T$ and convert the sum into the integral:
\begin{equation}
\rho^{xx}_{12} \simeq
- 0.0095 
\frac { \hbar }{e^2 n_1 n_2 d^4}
\frac{T}{E_0}
\left (\frac{a_B}{d} \right )^{3/2}.
\label{17}
\end{equation}

In region III ($\omega_c \gg \sqrt{E_0 \Delta}$)  the $j=1$ term 
dominates the sum. The corresponding expression for the plasmon drag
resistance is exponentially
small,
\begin{equation} 
\rho^{xx}_{12} \simeq - 2.7 \times 10^{-6} 
\frac { \hbar }{e^2 n_1 n_2 d^4}
\frac{(\hbar \omega_c)^9 T}{\Delta^{9/2} E_0^{11/2}}
\left (\frac{a_B}{d} \right )^{3/2}\exp(-0.28 (\hbar \omega_c)^2/\Delta E_0).
\label{qexp}
\end{equation}
 This is due to the fact that the value of $q$ which is needed to bring the
magnetoplasmon pole to the vicinity of the Landau level
is large compared to $1/d$. Similar situation occurs in the region IV,
with the exponential suppression being  due to low temperature,
\begin{equation} 
\rho^{xx}_{12} \simeq - 1.12 \times 10^{-5} 
\frac { \hbar }{e^2 n_1 n_2 d^4}
\frac{(\hbar \omega_c)^8}{\Delta^4 E_0^4}
\left (\frac{a_B}{d} \right )^{3/2}\exp(-\hbar \omega_c/T).
\label{omexp}
\end{equation} 

 We will see below
that even the exponentially suppressed magnetoplasmon contribution
can efficiently compete with the quasiparticle one.

Now we will estimate quasiparticle contribution 
in all three regions.
 
Let us first consider region I
$(\omega_c \ll T \ll E_0 \Delta /\omega_c)$. In this parameter interval
the characteristic values of $\hbar \omega \simeq T$  and $qd \simeq 1$.
It follows from Eqs.\ref{6},\ref{7}  that ${\rm Re} \chi = \nu$ and
${\rm Im} \chi \ll \nu$. Thus we are in the normal screening regime.
Using the fact that $V(q)\nu \gg 1$ (in other terms,
$qa_B
\ll 1$) we obtain from Eq.\ref{2}:
\begin{equation}
\rho^{xx}_{12} \simeq - 0.00725 \frac {\hbar}{e^2} \frac {1}{n_1 n_2 d^4} \frac{T^2
a_B^2}
{\hbar^2 v_{F1} v_{F2}} \frac{\omega_c}{\Delta}.
\label{9}
\end{equation}

This magnetotransresistance is bigger than the zero-field value
(Eq.\ref{1}) by a factor
of $\omega_c/\Delta \gg 1$. This is due to the discreteness of
the electron spectrum in the magnetic field when the density of states
within
the Landau level remarkably increases. 

Region II ($\omega_c \ll E_0 \Delta/\omega_c \ll T$).
Though the main contribution to the drag in this region is due to the
magnetoplasmon mechanism, it is instructive to give 
 an estimation  of  the quasiparticle contribution.
As in the region I, $q \simeq 1/d$.
 The characteristic frequency, however,  is restricted by
the
value $\omega_{cut} \sim E_0 \Delta/\omega_c $ which is much smaller than the
temperature
(though still
much larger than the cyclotron frequency). The
reason is the overscreening.
For estimations, we approximate ${\rm Im}\chi$ near the cyclotron resonance
by its value from Eq.\ref{6} at $qd \simeq 1$. That gives
${\rm Im}\chi  \simeq \nu \omega/\omega_{cut}$. A good estimation
for ${\rm Re}\chi$ near the resonace is $\nu$ for $\nu \gg {\rm Im}\chi$
and ${\rm Im}\chi$ otherwise. 
Thus the integrand in Eq.(\ref{2}),
$({\rm Im} \chi)^2  /\mid {\cal E} \mid^2 \propto
({\rm Im} \chi)^2/({\rm Re}^2 \chi + {\rm Im}^2 \chi)^2$
 achieves  a  maximum at $\omega_{cut}$.
Since the interval of integration over frequency is effectively 
smaller than the
temperature,
the transresistance exhibits a linear temperature dependence:
$\rho^{xx}_{12}\sim (\hbar /e^2) (1/n_1 n_2 d^4) (a_B^2/
d^2) (T /E_0)$. 

Let us  compare this with $B=0$ case. There, the linear temperature
dependence starts \cite{theory} at $T > E_0$.
This is because the absorbtion of
energy due to the Landau damping mechanism is possible only for frequencies
smaller than
$v_F /d$. Thermal frequencies  are ineffective, because
corresponding
phase velocities are larger than $v_F$.  
In the presence of a magnetic
field, the role of a cut-off energy
is taken  by a much smaller value $E_0 \Delta/\omega_c$. So that, the
linear temperature dependence of the transresistance starts at much
lower temperatures.

 Region III ($\sqrt{E_0 \Delta} \ll \hbar \omega_c \ll T$).
Here $\omega_{cut}$ becomes smaller than $\omega_c$.
There is overscreening at $qd \simeq 1$ for all cyclotron
resonances.
The contribution of resonances to  integral Eq.(\ref{2}) is of the order of
$\rho^{xx}_{12}\sim (\hbar /e^2) (1/n_1 n_2 d^4) (R_c^2 a_B^2/
d^4) (T \Delta^3/\omega_c^4)$.

The main contribution is determined by the frequencies
$\omega \sim \Delta \ll \omega_c$. So that the quasiparticles
are created within the same Landau level. For these frequencies
we always have the normal screening situation.
We can set ${\rm Re} \chi =\nu$. The ${\rm Im} \chi$ is determined by
the $j=0$ term of Eq.(\ref{6}).
We obtain 
\begin{equation}
\rho^{xx}_{12}=
- \frac {\hbar}{e^2} \frac {2}{\pi^4}\frac{1}{n_1 n_2 }
\int \frac{d^2q q^2}{(2\pi)^2}\frac{1}{\sinh^2(qd)}
\frac{T a_B^2}
{\Delta^2 R_{c1} R_{c2}}
\int_0^{2\Delta} \frac{d\omega}{2\pi}X_{im}^2 \left(\frac{\omega}{2\Delta}
\right)
\simeq - 0.0011\frac {\hbar}{e^2}
\frac {1}{n_1 n_2 d^4} \frac{\omega_c^2 T a_B^2}
{\Delta v_{F1} v_{F2}}
\label{10}
\end{equation}
It is interesting to note that this magnetic and temperature dependence
coincides precisely with the observed one \cite{Hill} if we assume that
$\Delta$ does not depend on magnetic field. Indeed, the magnetic dependence
of $\Delta$ is rather weak. These experiments
were performed in rather strong magnetic field where only few Landau
levels were occupied so that one should not expect quantitative
agreement with our calculations. From the other hand, the linear
T-dependence is rather remarkable. We believe that in any case
it indicates a reduction of typical excitation energies to values
much smaller than $T$, possibly due to overscreening at energies
of the order of $T$.

Now we are in position to compare quasiparticle and magnetoplasmon
contribution and thus to set the borders of the gray-shaded regions
in Fig. 1. The magnetoplasmon contribution dominates throughout the
region II. In the region I we compare expressions (\ref{16}) and
(\ref{9}). Magnetoplasmon contribution dominates provided
$T \simeq E_0 \Delta/\hbar\omega_c (a_B/d)^{1/4}$. Since experimentally  $d \simeq a_B$ this happens in fact close to the border
between the regions I and II $ T \simeq E_0 \Delta/\hbar\omega_c$. 
The exponentially small
magnetoplasmon 
contribution given by (\ref{qexp}) competes with (\ref{10}) in the region III
and dominates provided 
$0.28 \omega_c^2/(\Delta E_0) < 
\ln((\omega_c^2 /E_0 \Delta)^{7/2} 
(d/a_B)^{1/2})$. 
It also dominates in the region IV if $T/\hbar\omega_c > 1/\ln(\omega_c^4 T d^{1/2}/(E_0^2 \Delta^3 a_B^{1/2}))$. The latter condition is obtained 
by comparing expressions (\ref{omexp}) and (\ref{11}).

\section{Results for low temperatures}
In this section we present our results for low temperatures $T\ll \hbar
\omega_c$. Owing to energy limitations, only the states of the upper
partially filled Landau level are involved into the drag.
This makes the transresistance sensitive to concrete value of this
filling factor. In addition to slow dependence on magnetic field,
the drag effect exhibits oscillatory dependence on inverse magnetic
field related to the filling factor. The detailed study of these
{\it Shubnikov-de Haas} oscillations is beyond the limits
of the present work and will be presented elsewhere. This is why
we provide here analythical results for the regions IV and VI only.
As to the regions V and VII, we present below estimations of the transresistance
for typical filling factors rather than detailed analythical results.

 In the region IV ($\Delta \ll T \ll \hbar \omega_c
\ll \sqrt{TE_0}$) the situation is the most straightforward one
since it follows from Eq.(\ref{8}) that $Im \chi \ll \nu$. Thus
we encounter here the normal screening and we may set
 $Re \chi = \nu$. We obtain for the transresistance:
\begin{equation}
\rho^{xx}_{12}=
-0.0011
 \frac {\hbar}{e^2}
\frac {1}{n_1 n_2 d^4} \frac{ a_B^2}
{ v_{F1} v_{F2}} \frac{\omega_c^4}{T\Delta}
f_{n_1}(1-f_{n_1})
f_{n_2}(1-f_{n_2})
\label{11}
\end{equation}
$f_{n_{1,2}}$ being filling factors in the layers.

Even if $T \gg \Delta$, we encounter  overscreening in the region V 
($\sqrt{TE_0} \ll \hbar \omega_c \ll
E_0$). For estimations, we set ${\rm Im} \chi \sim {\rm Re} \chi \gg \nu$ 
and obtain
\begin{equation}
\rho^{xx}_{12} \sim (\hbar/e^2)
(1/n_1 n_2 d^4)( a_B^2  R_{c1} R_{c2}/d^4) (T^3/\omega_c^2 \Delta).
\label{12}
\end{equation}
As a consequence of the overscreening, 
the transresistance rapidly decreases with increasing magnetic field.

At low temperature $T \ll \Delta$ the integral in Eq. 1 is
contributed by $\omega \simeq T$. This gives rise to featureless
$T^2$ temperature dependence of Eq. \ref{1} with the coefficient
depending on magnetic field.
The region VI ($T \ll \Delta$, $\omega_c \ll\sqrt{E_0 \Delta}$) again corresponds to normal
screening. We take ${\rm Im}$ from Eq.(\ref{nultempchi}) and set
$Re \chi = \nu$. 
This yields
\begin{equation}
\rho^{xx}_{12}=
- 0.0031 \frac {\hbar}{e^2}
\frac {1}{n_1 n_2 d^4} \frac{ a_B^2}
{ v_{F1} v_{F2}} \frac{T^2 \omega_c^4}{\Delta^4}
(1-\frac{(\mu_1 - \epsilon_n)^2}{\Delta^2})(1-\frac{(\mu_2 - \epsilon_n)^2}{\Delta^2})
\end{equation}
where $\mu_{1,2}$ are chemical potentials in the layers.
They are related to filling factors $f_{n_1,n_2}$
by means of 
\begin{equation}
f_n = 1/2 + (1/\pi)[ (1-\frac{(\mu_1 - \epsilon_n)^2}{\Delta^2})
\sqrt{1-(1-\frac{(\mu_1 - \epsilon_n)^2}{\Delta^2})^2} + \arcsin (1-\frac{(\mu_1 - \epsilon_n)^2}{\Delta^2})].
\end{equation}
For a typical filling factor,
the effect 
 is bigger than zero filed transresistance by a factor of $(\omega_c/\Delta)^4$.

With increasing the magnetic field, we enter the region of overscreening
(VII, $T \ll \Delta$, $\omega_c \gg \sqrt{E_0 \Delta}$.) 
Again we set $ {\rm Re} \chi \gg \nu$ 
and obtain
\begin{equation}
\rho^{xx}_{12} \sim (\hbar/e^2)
(1/n_1 n_2 d^4)( a_B^2  R_{c1} R_{c2}/d^4) (T^2/\omega_c^2).
\label{posledneee}
\end{equation}
that decreases with increasing magnetic field.

\acknowledgments
\section*{}
This work is  part of the research program of the "Stichting voor
Fundamenteel
Onderzoek der Materie (FOM)", which is financially supported by the
"Nederlandse Organisatie voor Wetenschappelijk Onderzoek (NWO)".
We  thank  G. E. W. Bauer and L. I. Glazman for useful discussions.
A.V. Khaetskii acknowledges NWO for its  support of
his stay in Delft.
He is also grateful to F. Hekking for collaboration at the earlier stage of
the work, J.T. Nicholls and  D.E.
Khmelnitskii for useful discussions and 
M. Pepper for hospitality and financial
support at Cavendish Laboratory where a 
part of this work has been done.

\appendix
\section*{}
In this work we disregard rapidly oscillating parts of 
polarization function given by Eq. \ref{3}. This means we approximate
Bessel function squares in Eq.\ref{3} in the limit of $qR_c \gg 1$
as  $J_m^2(qR_c) \approx 1/\pi q R_c$ rather than as
\begin{equation}
J_m^2(qR_c) \approx \frac{1}{\pi q R_c} ( 1 + \cos(2qR_c -\pi m -\frac{\pi}{2})),  
\end{equation} 
which is the mathematically correct expression. Let us  explain why.

First let us note that if we take these oscillating parts into account
it would significantly alter our results. In the normal screening regime
it would give an extra factor of $3/2$, since the answer is proportional
to $({\rm Im} \chi)^2$. The answer would change even more drastically in
the overscreening regime. The point is that oscillating terms would
set the polarization function to (almost) zero for $q$ corresponding
to zeros of the Bessel function. No overscreening would occur near
these points and their close vicinity would dominate the drag.

All this would lead to very sophisticated and extremely unstable
picture of the drag effect. Fortunately, we are able to present some
arguments that allow to disregard oscillating terms in polarization
function.

The physical origin of the oscillating terms can be best understood
in the language of semiclassical electron trajectories in magnetic
field. A classical
trajectrory can not move from the starting point further
than $2 R_c$. This is why the polarization function in coordinate 
representation, $\chi(x,x')$ has a sharp edge at $\mid x-x'\mid = 2R_c$.
This gives rise to Fourier components $\simeq \cos(2q R_c)$. 

If the edge is not sharp, the oscillating part is exponentially
suppressed. The suppression is of the order of $\exp(-q^2 (\delta R)^2)$,
$\delta R$ being a typical rounding of the edge.
By virtue $\delta R$ is the typical uncertainty of the coordinate
of the electron that makes a half of Larmor circle.
Such an uncertainty can be of quantum-mechanical origin.
In this case we estimate $\delta R \simeq \sqrt{R_c/k_F} \ll R_c$.
Another cause of uncertaincy may be small-angle scattering by
smooth potential fluctuations in the heterostructure.
For this case we estimate $\delta R \simeq \sqrt{R_c^3/l_{sa}}$, 
where transport mean free path $l_{sa} \gg
R_c$. It is interesting to note that scattering on point-like
defects does not contribute to $\delta R$ provided $\omega_c \tau \gg
1$.

Now we note that typical $q$ contributing to the drag resistance
are  of the order of $1/d$. We conclude that the oscillating part 
is exponentially suppressed provided
$d < \sqrt{R_c/k_F}$ or $d < R_c \sqrt{R_c/l_{sa}}$.
We assume that at least one of these conditions is fulfilled.

\begin{figure}
\caption{Seven regions of different analytical behaviour of the
transresistance in the intermediate magnetic field regime.
Note log-log scale. First number in each region corresponds to temperature
exponent, second number indicates magnetic field exponent.
Vertical line corresponds to the condition $(\hbar\omega_c)^2 \simeq E_0 \Delta$.}
\label{fig1}
\end{figure}

\begin{figure}
\caption{The functions $X_im(x)$ and $X_re(x)$ determine
the shape of ${\rm Im} \chi$, ${\rm Re} \chi$ in the
vicinity of each cyclotrone resonance.}
\label{fig2}
\end{figure}

\begin{figure}
\caption{Electron adsorption in $\omega,q$ plane. Left plane:
electron adsorption occurs i. in narrow strips around $\omega=0$ and
cyclotron resonances (particle-hole continium) ii. on the magnetoplasmon
dispersion curves. Right plane: intersection of the
magnetoplasmon dispersion curve and the edge of the strip at a smaller
scale. We illustrate splitting of the magnetoplasmons and the level width
$\Gamma$ they acquire inside the strip.}
\label{fig3}
\end{figure}

\begin{references}

\bibitem{Gramila}
T.~J. Gramila, J.~P. Eisenstein, A.~H. MacDonald, L.~N. Pfeiffer,
and K.~W. West, Phys.Rev.Lett. {\bf 66}, 1216 (1991);
T.~J. Gramila, J.~P. Eisenstein, A.~H. MacDonald, L.~N. Pfeiffer,
and K.~W. West, Surf.Sci. {\bf 263}, 446 (1992);
T.~J. Gramila, J.~P. Eisenstein, A.~H. MacDonald, L.~N. Pfeiffer,
and K.~W. West, Phys.Rev.B. {\bf 47}, 12957 (1993);
 U. Sivan, P.M. Solomon, H. Shtrikman, Phys. Rev. Lett. {\bf 68}, 1196
(1992).

\bibitem{theory}

A.-P. Jauho, H. Smith, Phys. Rev. B {\bf 47}, 4420 (1993); L. Zheng, A.H.
MacDonald, Phys. Rev. B
{\bf 48}, 8203 (1993);
A. Kamenev, Y. Oreg, PRB {\bf 52}, 7516 (1995);
K. Flensberg, B.Y.-K. Hu, Antti-Pekka Jauho and J. Kinaret,  Phys. Rev. B {\bf
52}, 14761
(1995).

\bibitem{Flensberg}

K. Flensberg, B.Y.-K. Hu, Phys. Rev. Lett. {\bf 73}, 3572 (1994);
K. Flensberg, B.Y.-K. Hu, Phys. Rev. B {\bf 52}, 14796 (1995).

\bibitem{Hill}
N.P.R. Hill, J.~T. Nicholls, E.~H. Linfield, M. Pepper, 
D.~A. Ritchie, A.~R. Hamilton, and G.~A.~C. Jones, 
J.Phys.: Cond.Matt. {\bf 8}, L557 (1996),


\bibitem{Rubel}
H. Rubel,A. Fischer, W. Dietsche, K. von Klitzing, and K. Eberl, Phys. Rev. Lett {\bf 78}, 1763 (1997);
J.P. Eisenstein,L. N. Pfeiffer, and K. W. West, Bull. Am. Phys. Soc. {\bf 42}, 486 (1997);
N.P.R. Hill et al., unpublished, to
appear in proceedings of EP2DS 12 (Tokyo, 1997).

\bibitem{Bonsager}
M.C. Bonsager, K. Flensberg, B. Y.-K. Hu, A.-P. Jauho, 
Phys. Rev. Lett. {\bf 77}, 1366 (1996).
\bibitem{Bonsager2}
M.C. Bonsager, K. Flensberg, B. Y.-K. Hu, A.-P. Jauho,
 Phys. Rev. B {\bf 56}, 10314 (1997).
\bibitem{Ando}
T. Ando and Y. Uemura, J. Phys. Soc. Jpn. {\bf 36}, 959 (1974);
T. Ando, A. B. Flower, and F. Stern, Rev. Mod. Phys., {\bf 54}, 437 (1982). 

\bibitem{Aleiner}
I. L. Aleiner and L. S. Glazman, Phys. Rev. B {\bf 52}, 11296 (1995).
\bibitem{Old}
N. J. M. Horing and M. M. Yildiz, Ann. Phys. (NY), {\bf 97}, 216 (1976).

\bibitem{Magnetoplasmons}
C. Kallin and B. I. Halperin, Phys. Rev. B {\bf 30}, 5655 (1984);
D. Antoniou and A. H. MacDonald, Phys. Rev. B {\bf 46}, 15225 (1992).

\bibitem{Sarma}
S.D. Sarma and A. Madhukar, Phys. Rev. B {\bf 23}, 805 (1981).


\end{references}
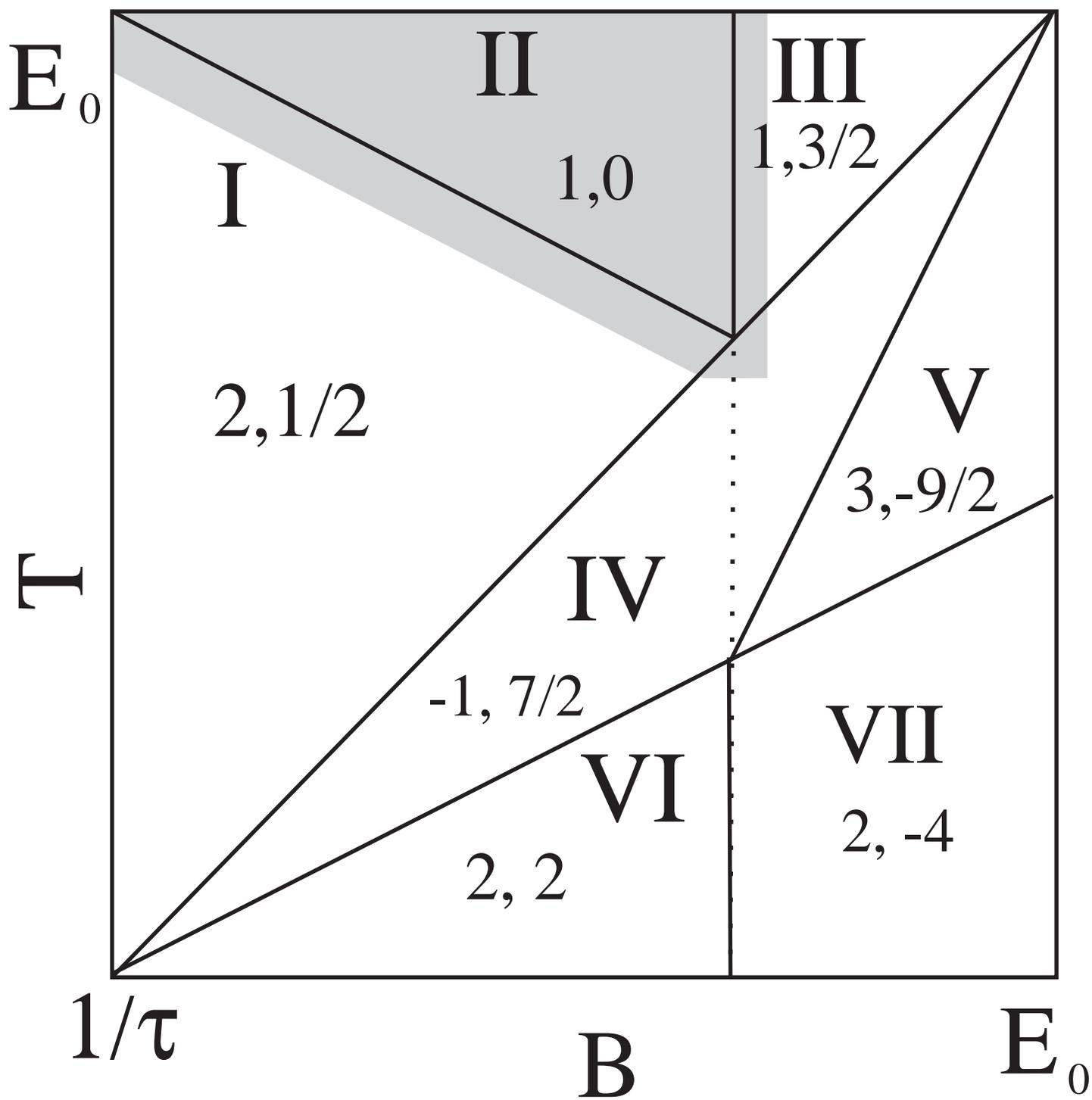
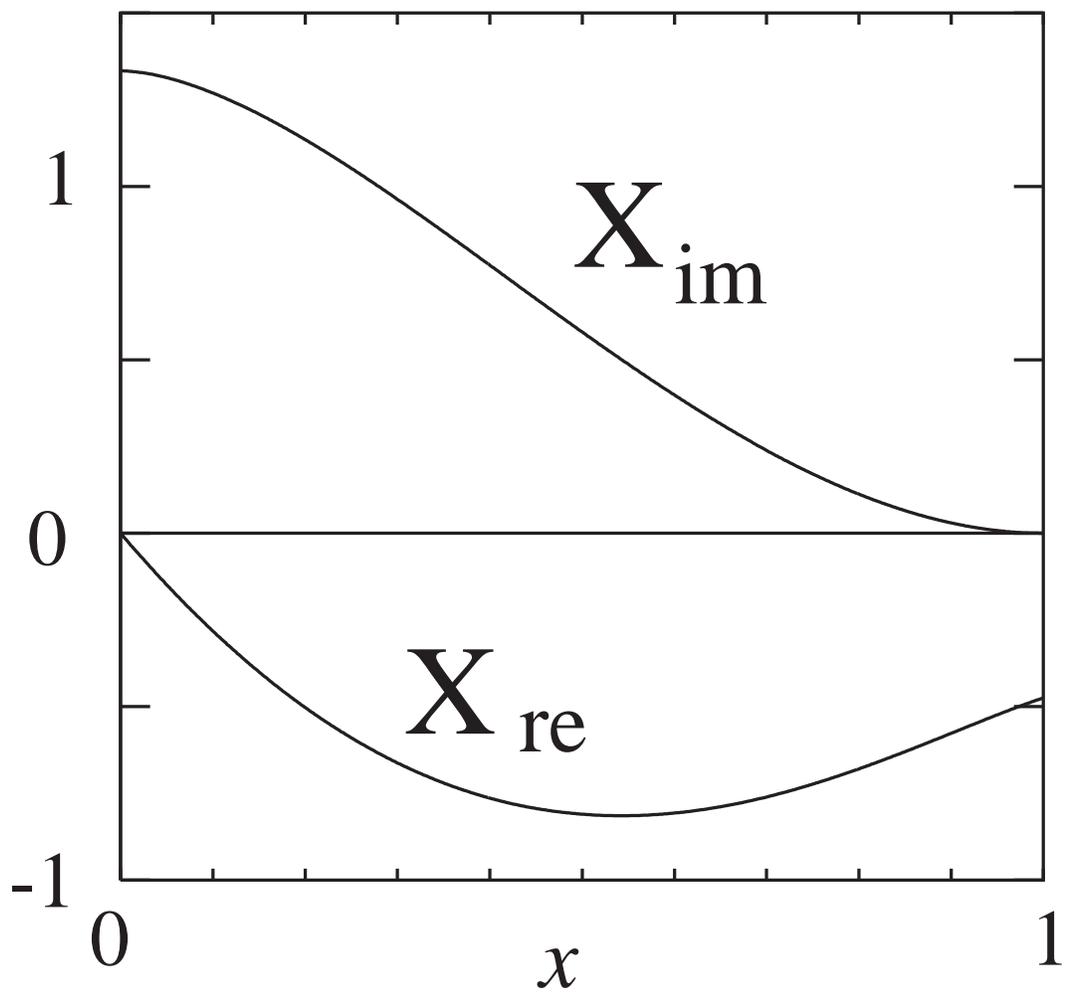
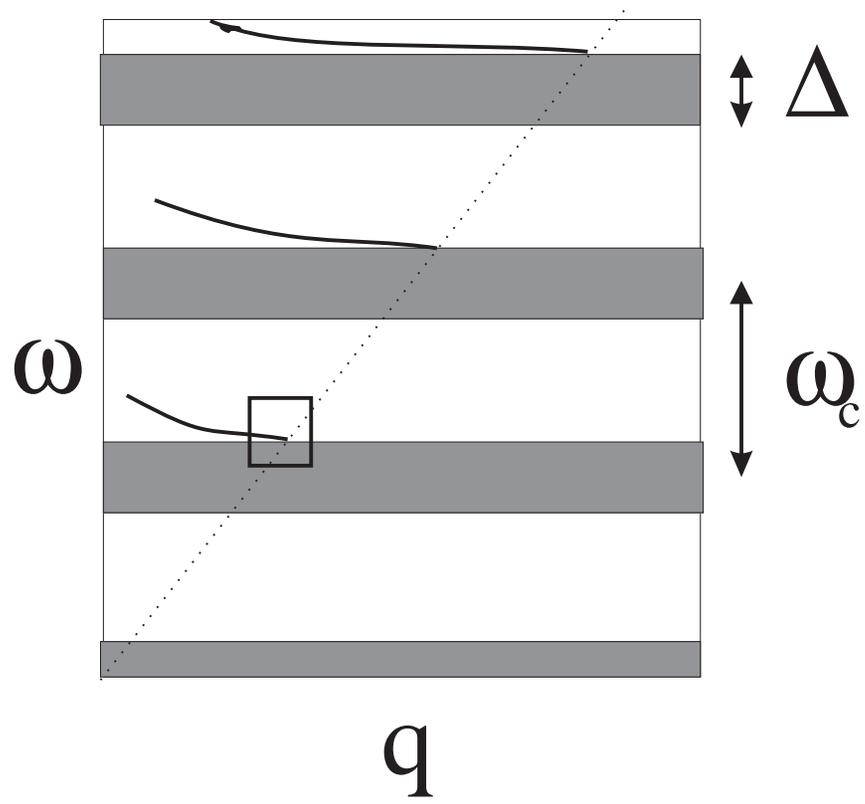
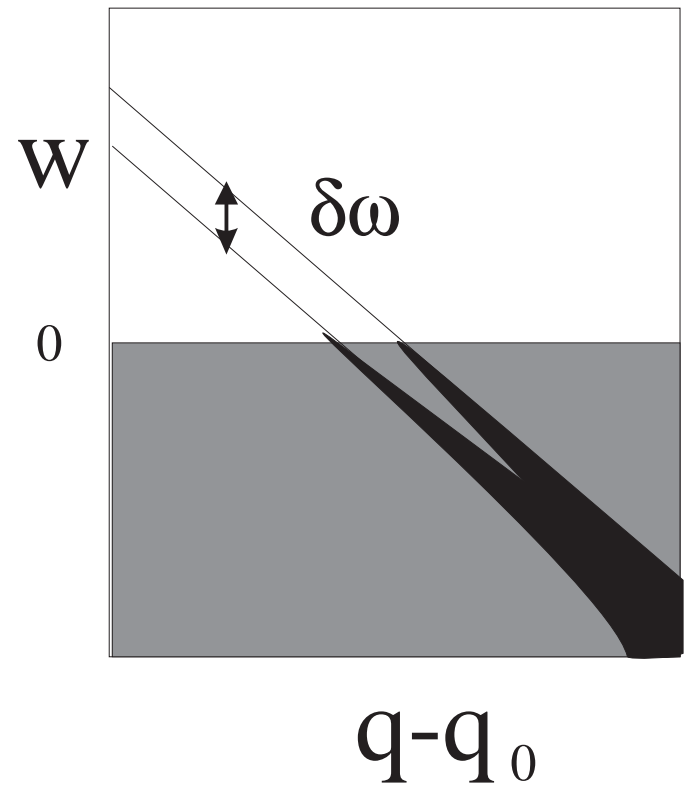
\end{document}